# CHAPTER 4

# Millions of Co-purchases and Reviews Reveal the Spread of Polarization and Lifestyle Politics across Online Markets

Alexander Ruch[1], Ari Decter-Frain[2], Raghav Batra[3]

[1] Spotify `aruch@spotify.com`; Cornell University, `amr442@cornell.edu`
[2] Cornell University, `agd75@cornell.edu`
[3] Cornell University, `rb698@cornell.edu`

**Abstract**

Polarization in America has reached a high point as markets are also becoming polarized. Existing research, however, focuses on specific market segments and products and has not evaluated this trend's full breadth. If such fault lines do spread into other segments that are not explicitly political, it would indicate the presence of lifestyle politics – when ideas and behaviors not inherently political become politically aligned through their connections with explicitly political things. We study the pervasiveness of polarization and lifestyle politics over different product segments in a diverse market and test the extent to which consumer- and platform-level network effects and morality may explain lifestyle politics. Specifically, using graph and language data from Amazon (82.5M reviews of 9.5M products and product and category metadata from 1996–2014), we sample 234.6 million relations among 21.8 million market entities to find product categories that are most politically relevant, aligned, and polarized. We then extract moral values present in reviews' text and use these data and other reviewer-, product-, and category-level data to test whether individual- and platform-level network factors explain lifestyle politics better than products' implicit morality. We find pervasive lifestyle politics. Cultural products are 4 times more polarized than any other segment, products' political attributes have up to 3.7 times larger associations with lifestyle politics than author-level covariates, and morality has statistically significant but relatively small correlations with lifestyle politics. Examining lifestyle politics in these contexts helps us better understand the extent and root of partisan differences, why Americans may be so polarized, and how this polarization affects market systems.

**Keywords:** markets, networks, polarization, lifestyle politics, morality.

## 1 Introduction

Americans are more polarized today than they have been at any other point in recent history (Finkel et al. 2020). Though partisan divisions on moral issues have been closing over time (Baldassarri and Park 2020), between-group animosity has steadily grown since the 1970s (Boxell, Gentzkow, and Shapiro 2020; Iyengar et al. 2019; Iyengar, Sood, and Lelkes 2012). Belief consolidation may also contribute to rising polarization, whereby the beliefs by which partisans sort themselves into opposing groups have become more clustered into tightly coupled "packages" with greater in-group alignment over the last 44 years (DellaPosta 2020). An example of this process, the author shows, is attitudes toward taxation becoming more coupled with opinions on climate change, which then become increasingly entwined with religious beliefs – spurring a more polarized ideological network through which partisans are clearly divided.

These polarization dynamics are also apparent in markets and science. Though conservatives and liberals both purchase scientific literature, they consume different kinds of scientific books over different fields' market segments (F. Shi et al. 2017). For example, conservatives tend to buy books that are more peripheral and applied in market networks compared to liberals, who buy a broader range of scientific books. Overall, polarization among the co-purchases of scientific and political products is almost three times greater than the polarization that exists among co-purchases of non-scientific and political products.

While this study covered substantial breadth in examining partisan differences in consumption – including not only broad fields like physical sciences, life sciences, social sciences, and humanities, but also a wide range of subfields within these disciplines of study – it remains to be seen if these divisions also extend beyond science to other market segments. Lifestyle politics (DellaPosta, Shi, and Macy 2015) like this develop when ideas and behaviors that are not inherently political, including support for guns or gender rights, become politically aligned through their direct and indirect relationships with explicitly political things, like political media/policies and politicized events (e.g., corporate action on social issues). For example, after Levi Strauss & Co. pledged over $1 million to support ending gun violence and strengthening gun control laws, the jean company became progressively aligned with liberals while conservatives aligned themselves more with Wrangler (Kapner and Chinni 2019).

The traces of lifestyle politics are pervasive. For example, analyses of Twitter co-following show the stereotypes of "Tesla liberals" and "bird hunting conservatives" have empirical support (Y. Shi et al. 2017). It is less clear, however, what explains these dynamics. DellaPosta et al. (2015) argue that network autocorrelation stemming from homophily and social influence can generate lifestyle politics in absence of psychological variables and other social factors such as demographics. Social psychologists on the other hand argue that different moral attributes implicitly inherent to lifestyles, thoughts, and behaviors appeal more strongly to different ideological groups, especially regarding "culture war attitudes" (Graham et al. 2013; Koleva et al. 2012). For example, moral values like care/harm and fairness/cheating have greater appeal to liberals, whereas conservatives support all moral values highly (Graham, Haidt, and Nosek 2009).

We contribute to this research by evaluating the pervasiveness of polarization and lifestyle politics across different product segments in large markets and by assessing the extent to which consumer- and platform-level network effects and moral values may explain lifestyle politics. Specifically, using graph and language data from nearly 20 years of product and review data from Amazon.com (82.5M reviews of 9.5M products in addition to product and category metadata from 1996–2014), we sample 234.6 million observations of relations among 21.8 million market entities to assess the product categories that are most politically relevant, aligned, and polarized. We then extract moral values present in reviews' text and use these data and other reviewer-, product-, and category-level data to test whether individual- and platform-level network factors explain lifestyle politics better than products' implicit moral attributes. Understanding lifestyle politics in these contexts can help us to better understand the extent and root of partisan differences and why Americans are polarized.

## 2 Data and Sampling

### 2.1 Market Data Sampling and Network Construction

Market network data for this study came from He and McAuley's (2016) publicly available sample of aggressively deduplicated Amazon.com reviewer, product, and metadata data[1]. These data are free from duplicate reviews by users and include metadata on products' descriptions, price, sales-rank, brand info, and co-purchasing links to other products (i.e., products bought together, also bought later, bought after viewing, and also viewed – which are platform-level recommendations). Hand-labeled and verified conservative and liberal book titles from Shi and colleagues' (2017) study of political and scientific polarization on Amazon were extracted from a local server repository[2]. The market network and political seed data were inserted into a PostgreSQL database, after which political products and market network data were matched using the partial ratio of the Levenshtein distance between seeds' titles and products' titles in the database[3]. All matches were evaluated by two members of the research team to assure labels were validly applied to the market data. In some cases, duplicate labels were found among and removed from the political seed data, including cases when a political product existed in multiple formats (e.g., softcover, hardcover, e-book, audio), because the deduplicated market network data included only one format of each title.

This process generated 1,116 labels for liberal and conservative books in the market network data[4]. From these seeds, two waves of a two-step breadth-first search network sampling were performed. The first step of the first wave focused on sampling related products and sampled from the seed products to their distinct co-purchased products ($n = 8,944$) and then sampled all reviewers of these products ($n = 552,853$). The second step of the first wave sample focused on sampling reviewers of seed products and sampled all distinct reviewers of the seed products ($n = 66,816$) and sampled all distinct co-purchases related to these products ($n = 1,891,249$). After combining the results of the two steps of the first wave sample and removing duplicate information, product-level metadata (e.g., brand and category) information were sampled from products ($n = 724,532$), which yielded a total of 2,319,244 observations for the first wave sample. For the second wave, the same sampling procedures were repeated out from the first wave data, which yielded 47,712,826 distinct reviews for 1,871,264 distinct products. Overall, this process extracted approximately 33% of the reviews and 20% of the products from the original sample collected by He and McAuley (2016). **Appendix A** describes the different kinds of information retrieved from this sampling process and others that are noted in the following sections.

A heterogeneous network was constructed from these data using `graph-tool`, a high-performance Python library for graph modeling (Peixoto 2014). This type of network differs from that of Shi et al. (2017) in important ways. First, they studied a one-mode projection of a bipartite network in which the graph only included product nodes and product-to-product edges from co-purchases. The heterogeneous network, by contrast, has nodes representing multiple entity types (e.g., authors, products, brands, categories) and relations (e.g., author-product edges for reviews, product-product representing different kinds of co-purchases, etc.). Heterogeneous graphs not only provide more information on networks, but they are also more flexible and allow for more kinds of relationships to exist between entities. For example, in a bipartite projection, one product can only be related to another product through a co-purchase; however, in a heterogeneous graph, one product can relate to another product through a co-purchase or the platform's recommendation, co-review (i.e., users' behavior), and co-membership in the same brand or category (i.e., market segment). Having more paths of connection, therefore, allows heterogeneous graphs to discover more possible connections between products that may be missing from bipartite networks. See **Appendix B** to view the degree distribution for the heterogeneous graph across different edge types.

---

[1] These data are available at http://jmcauley.ucsd.edu/data/amazon/links.html. Note that this work did not meet the definition of human subject research from Cornell University's Institutional Review Board for Human Participants (https://researchservices.cornell.edu/compliance/human-research), as when data were collected they were extracted from an existing database that was openly available in the public domain.

[2] A similar seed set is available in the paper's supplemental information: https://www.nature.com/articles/s41562-017-0079#Sec14

[3] Python's `fuzzywuzzy` library was used to perform these calculations: https://github.com/seatgeek/fuzzywuzzy.

[4] We could not match 155 of the 1,446 political seeds from Shi et al. (2017a) with the market network products, and we removed 175 duplicates from the remaining 1,291 products, which lead the number of seed political books in this study to be slightly less than that of Shi et al. (2017a), who had 1,256 political seeds. Unmatched products likely exist because the Shi et al. (2017a) data are newer than the market network data and because the market network dataset is not a complete census of Amazon products, which is apparent because not all products that exist as co-purchases in He and McAuley's (2016) metadata table have matching rows in their product table.

The full heterogeneous network created from the two sampling waves included 15,255,656 author nodes, 6,485,395 product nodes, 85,642 brand nodes, and 15,798 category nodes – which totaled to approximately 21.8 million distinct entities[5]. Over 151.7 million edges existed between entities, all of which existed within a fully-connected component. To remove poorly-connected nodes from the graph, 5-core decomposition was performed. The resulting 5-core graph had 6,170,353 nodes (15,672,139 fewer than the full graph), 3,968,438 of which were products, and 126,362,653 edges (25,347,594 less than the full graph). To reduce the graph further for analyses of the most popular products and most active reviewers, 20-core decomposition reduced the heterogeneous network to 2,032,620 nodes and 89,347,563 edges. Finally, co-purchase edges were added between products that were co-reviewed by users but which had no existing co-purchase edges, which increased the number of edges in the graph to 234.6 million – a 2.6-fold increase, again demonstrating the ability of heterogeneous graph to discover complex and hidden relationships between entities. The variety of relation types present in the heterogeneous network grants one an opportunity to test hypotheses that are untestable with bipartite data. For example, one can test different effects from individuals (e.g., selection effects from users' past behavior) versus platform mechanics (e.g., filter bubbles stemming from recommendation engine suggestions).

## 2.2 Morality Data

Hoover et al. (2020) curated a set of approximately 35 thousand tweets, each of which was labeled by multiple annotators for the presence of none, one, or more moral values (i.e., a multi-label task). Given the ids for these tweets, we were able to extract 23,455 tweets from the Twitter API[6]. Labels for these tweets' text included non-moral ($n = 17{,}370$), care ($n = 4{,}490$), harm ($n = 5{,}916$), fairness ($n = 3{,}858$), cheating ($n = 4{,}756$), loyalty ($n = 4{,}120$), betrayal ($n = 2{,}779$), authority ($n = 4{,}055$), subversion ($n = 3{,}257$), purity ($n = 2{,}029$), and degradation ($n = 2{,}129$).

## 2.3 Text Processing

---

[5] To prove that that heterogeneous sampling approach discovers more relationships between products than can be captured with bipartite sampling, we simulated a bipartite sampling procedure from these data over product-product edges only, which yielded a sample of 344,621 products – 18.8 times fewer products than the heterogeneous sample. This difference is likely attributable to filtering effects from Amazon's recommendation engine.

[6] Some tweets labeled by Hoover et al. (2020) had been removed, deleted, or made private and were unavailable.

Text from tweets and Amazon reviews was preprocessed to remove emojis, @ signs, # signs for hashtags, "RT" retweet tokens, punctuation, URLs, newline symbols, extra spaces between words and at the start/end of text, stopwords, digits, and special characters encoded with an ampersand. All tokens (words) beyond 512 tokens were removed, as the transformer model that will classify the texts is limited to 512 tokens. Lastly, Tweets shorter than 5 tokens were removed from analysis as were Amazon reviews that were less than 30 tokens. Removing "short" Tweets left 21,232 texts for language models. Removing "short" Amazon reviews left 1,195,268 texts for classification.

# 3 Methods and Results

## 3.1 Product-level Political Classification

A semi-supervised relational graph convolutional network (RGCN; Kipf and Welling 2017) was used to expand the dataset of labeled political products and to classify the political alignment of products' network neighbors. With a small set of labeled examples (1,116 liberal and conservative products), the RGCN not only learns to classify products' political alignment but also that of those products' immediate neighbors (e.g., direct co-purchases), neighbors of neighbors (e.g., co-purchases of co-purchases), and neighbors of neighbors' neighbors through heterogeneous label propagation. This quality is useful in complex network like markets, as the RGCN can learn the political alignment of products through their indirect and distant relations with other products (e.g., that camouflage is likely conservative, as it is co-reviewed with military books, which are co-purchased with books by conservative politicians).

To increase the number of labeled political products, a RGCN was first trained on the original set of labeled political products from Shi et al. (2017) with the objective of correctly classifying their political alignment. Training, validation, and test sets were created from 80/10/10 splits across the 1,116 labeled products. For computational efficiency and because these data represent the core of the market network that is most closely tied to political products, the model only learned to classify political alignment for products from the first sampling wave's data across products and authors – excluding brand and category entities and relations[7]. The RGCN was constructed and trained using Python's `dgl` library for graph neural networks (Zheng et al. 2020), the hyperparameters[8] of which were optimized automatically with Bayesian optimization (via Tree-structured Parzen Estimators) using `optuna` and model tracking/logging with `MLflow` to discover the best fitting model.

After optimization, the RGCN's test set accuracy was 99.13% (cross entropy loss = 0.0551). Since the model trained on data labeled only for liberal and conservative products, the trained model can only classify other products as liberal or conservative and not as non-political. For example, of the unlabeled products, the RGCN classified 51.8% as being conservative and 48.2% as liberal, which is clearly wrong as not everything in a market is political. Therefore, this testing accuracy needs to be qualitatively validated against classifications the model makes of products outside the labeled dataset and additional labels for non-political products must be learned as well. To do this, title and classification data were extracted from products that the model had classified as leaning strongly liberal, strongly conservatively, and ambiguously (e.g., equally liberal and conservative). Two members of the research team inspected the products to validate their alignment, after which the newly validated labels were added to the dataset of labeled political products – totaling 1,093 conservative, 1,446 liberal, and 3,039 non-political labels after merging.

---

[7] Attempting to model the full data sample of all entity and relation types across both sampling waves exceeded the GPU's memory (11 GB on a NVIDIA 1080 Ti). The remaining products and relations used for training include the following canonical tuples: (author, reviews, product), (product, reviewed-by, author), (product, related-to, product), where the "related-to" relation represents any kind of co-purchase edge. All relations are modeled as undirected.

[8] Hyperparameters included training epochs, hidden layers, hidden nodes, learning rate, gradient norming/clipping, L2 regularization, and dropout probability.

This process of human-in-the-loop machine learning was performed for five iterations, after which validation yielded few payoffs. The final RGCN[9], which was trained on 1,355 conservative, 1,664 liberal, and 4,643 non-political labels collected from past validation iterations, had a test accuracy of 86.54% (cross entropy loss = 1.0545). Among the unlabeled products, this model classified 8.9% as leaning conservatively, 2.5% as leaning liberally, and 88.6% as being non-political – proportion cutoffs that have greater face validity than the first iteration of the RGCN, which classified nothing as non-political when most products should have been labeled as such. **Appendix C** shows figures representing the relationship between the first, second, and final RGCNs' decision thresholds and the proportion of labels accepted. The main takeaway from these figures is that iteratively training and providing the RGCN with more labeled examples leads the decision curve to go from flat and confident in the first RGCN to curved but still steep in the second RGCN to curved and tapered in the fifth and final RGCN – indicating that the model is becoming increasingly uncertain about the classifications it makes about many products, which is what one would expect from a model about the political alignment of products in a market where some products are very clearly political but where most are either tangentially political or clearly non-political.

Labels for liberal ($n = 22,358$) and conservative ($n = 34,196$) products with probabilities equal to or greater than 95% were merged into the heterogeneous network. **Figure 1** shows the 5-core graph with annotations for categories[10] and labeled/classified political products and reviews. As expected, clear clustering exists among products from different categories. Culture category products have the greatest engagement with political products, which is to be expected given that many explicitly political products are books. Entertainment-related products also have visible political overlap via their relationships with political movies and TV, sports and outdoor activities, and games. **Figure 2** shows the extent of polarization among political products, with liberal and conservative products divided clearly in half with little overlap and crossover across the dense political core.

---

[9] The final RGCN had 3 layers (i.e., it performed label propagation across products' neighbors, those neighbors' neighbors, and the neighbors of those 2-step neighbors), 19 hidden units at each layer, dropout of 68% (i.e., at each batch, 68% of nodes were deactivated to prevent overfitting), a learning rate of 0.05, gradient clipping at 3.358, L2 normalization with 1.66E-7, and 100 training epochs. Leaky ReLU activation was used at each layer of the model.

[10] **Appendix D** lists categories in the "main" and "big" category sets and notes how many products belong to each.

**Figure 1: Heterogeneous Market Network with Category/Political Annotations**

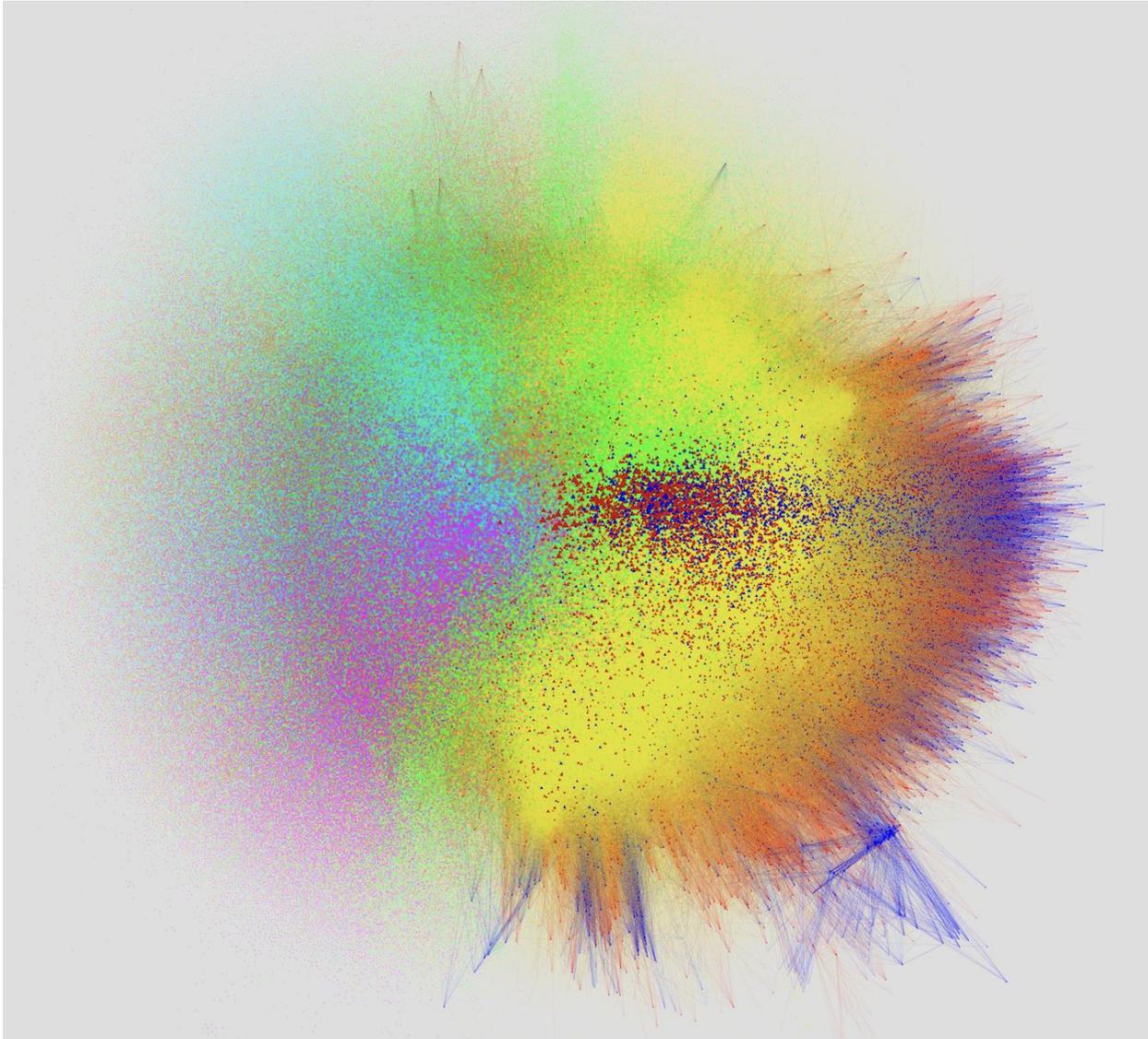

Note: Spring force-directed projection of the 5-core graph color-annotated with categories and political products/reviews: **red** = conservative; **blue** = liberal; **teal** = home; **orange** = products; **magenta** = personal/family; **yellow** = culture; **green** = entertainment; nodes: 5,505,655; edges: 120,322,391.

**Figure 2: Political Market Network Projection with Political Annotations**

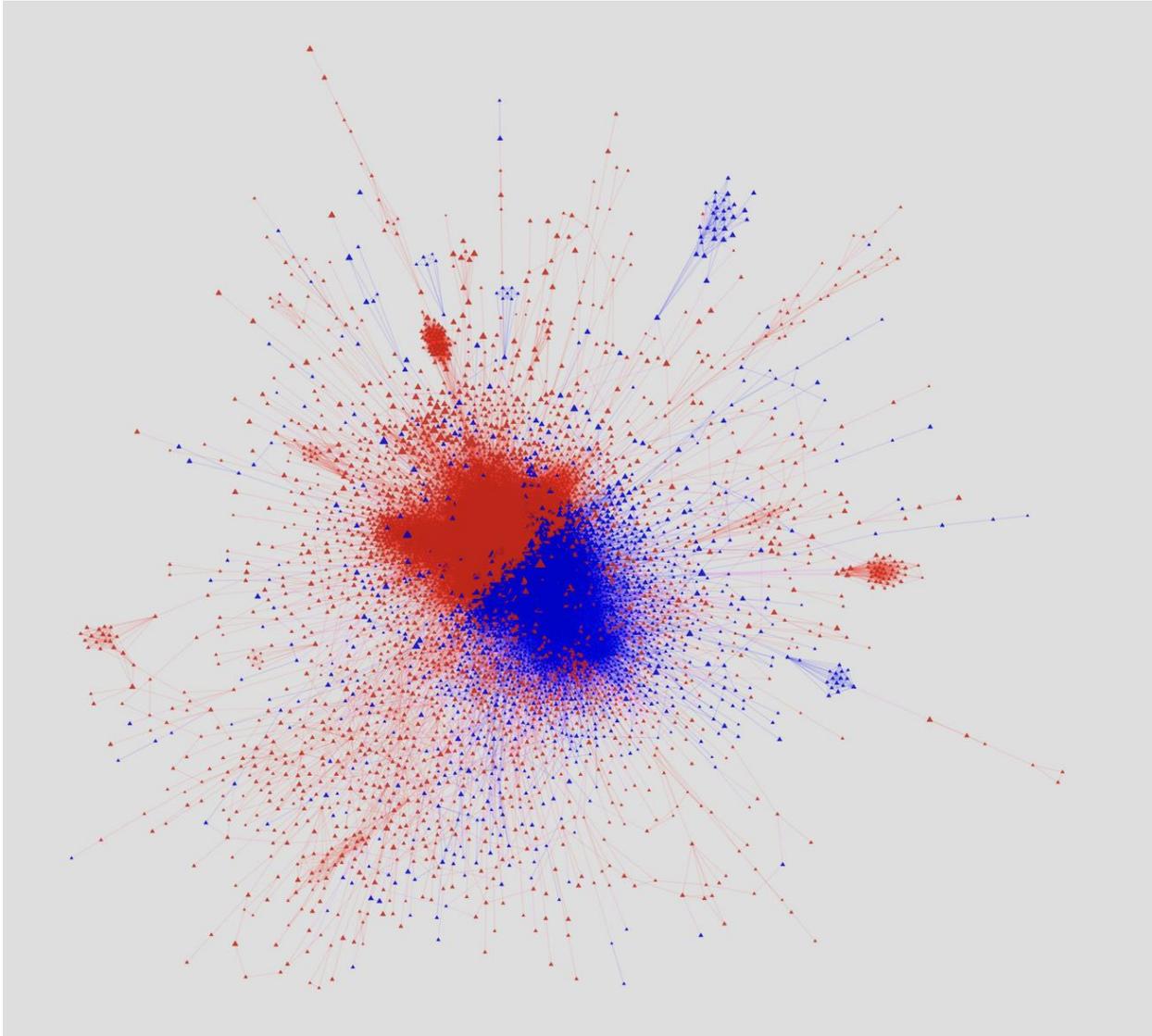

Note: Spring force-directed projection of the labeled/classified political graph color-annotated for political products/reviews: **red** = conservative, **blue** = liberal; nodes: 9,252, edges: 158,999.

## 3.2 Category-level Political Relevance, Alignment, and Polarization

To quantitatively measure the political relevance, alignment, and polarization of different market segments, we implemented the respective Bayesian measures of Shi et al. (2017) across the "big" and "main" product categories. The details of these measures are covered in the authors' Methods section; however, we express them below as well. In brief, political relevance measures how many edges for a product are tied to political products, while political alignment measures how many of those edges to political products are tied to conservative products. Political polarization measures how much greater the expected overlap between liberal and conservative products is in a category (i.e., the mean number of products with edges to liberal and conservative products after randomly shuffling products' edges repeatedly) compared to their observed overlap (i.e., how many products really have edges to both liberal and conservative products) relative to the variance of the sampling distribution for randomly shuffled edges (i.e., the expected overlap's variation from the repeated sampling process mentioned above). **Equation Set 1** describe these measures mathematically.

**Equation Set 1: Political Relevance, Alignment, Polarization Measures from Shi et al. (2017)**

$$Relevance = E[\theta|X] = \frac{X + \frac{dk_{political}}{m}}{K + d}$$

$$Alignment = E[\theta|X_{red}] = \frac{X_{red} + \frac{dk_{red}}{k_{red} + k_{blue}}}{K_p + d}$$

$$Polarization = \frac{E[O] - o}{\sqrt{var[O]}}$$

*Parameter Definitions and Values (when constant, for 20-core graph):*
- $\theta$ = estimated parameter (e.g., political relevance or political alignment)
- $X$ = observed edges between category products and political products
- $K$ = total edges attached to a category
- $d$ = prior strength (the average number of edges to political books over all categories): 302.3
- $k_{political}$ = total edges attached to political products: 2,796,590
- $m$ = total edges across categories: 212,929,627
- $K_p$ = total edges from a category to political products
- $X_{red}$ = total edges from a category to conservative products
- $k_{red}, k_{blue}$ = total links attached to conservative (1,818,415) and liberal products (978,175)
- $o$ = observed overlap (number of products in a category with liberal and conservative edges)
- $E[O]$ = expected overlap of political links in a category when randomly shuffled repeatedly
- $var[O]$ = variance of the sampling distribution of repeatedly randomly shuffled edges

Each of these equations was applied across all the "big" and "main" categories in the 20-core graph. **Table 1** and **Figure 3** show the magnitude of political relevance, alignment, and polarization over these categories. **Table 1** shows that cultural products are the most relevant and polarized products among the "big" categories, with 1% of these their edges going to political products and with 4.18 times more polarization than the second most polarized "big" category. Home products have the most alignment, with 60% of their political edges going to conservative products. **Figure 3** breaks these trends down further. Among the "main" categories, books have greatest relevance and by far the most polarization. Automotive products have the strongest alignment at 0.6694. By contrast, products in the music category lean most liberally with an alignment of 0.4895. The subgraphs for different categories visualized in **Figure 3** show how these political effects unfold over the graph. Evidence of books' relevance and polarization is made clear by the many political edges encircling the network with numerous clusters of red and blue edges spanning the category in small clusters. By contrast, the automotive category has far fewer connections to political products; however, the edges it does have are more oriented toward conservative products.

We validated these measures against two smaller subsets of products that are commonly polarized among partisans. As expected, products related to guns have a relevance of 0.006 (nearly the same as music & TV), an alignment of 0.646, and a polarization value of 45.90. Gender-related products have a relevance of 0.04 (four times more than books), an alignment of 0.251 (very liberal), and a polarization of 32.88. Therefore, we can be confident these measures are reliably representing the political relevance, alignment, and polarization of products and categories in the market network.

Table 1: Political Relevance, Alignment, Polarization Values for "Big" Categories

| Category | Relevance | Alignment | Polarization |
|---:|---:|---:|---:|
| Culture | 0.0104 | 0.4897 | 989.83 |
| Entertainment | 0.0037 | 0.5321 | 236.82 |
| Home | 0.0021 | 0.6012 | 137.07 |
| Personal & Family | 0.0021 | 0.5874 | 109.66 |
| Products | 0.0029 | 0.5952 | 150.42 |

Figure 3: Political Relevance, Alignment, Polarization Values for "Main" Categories

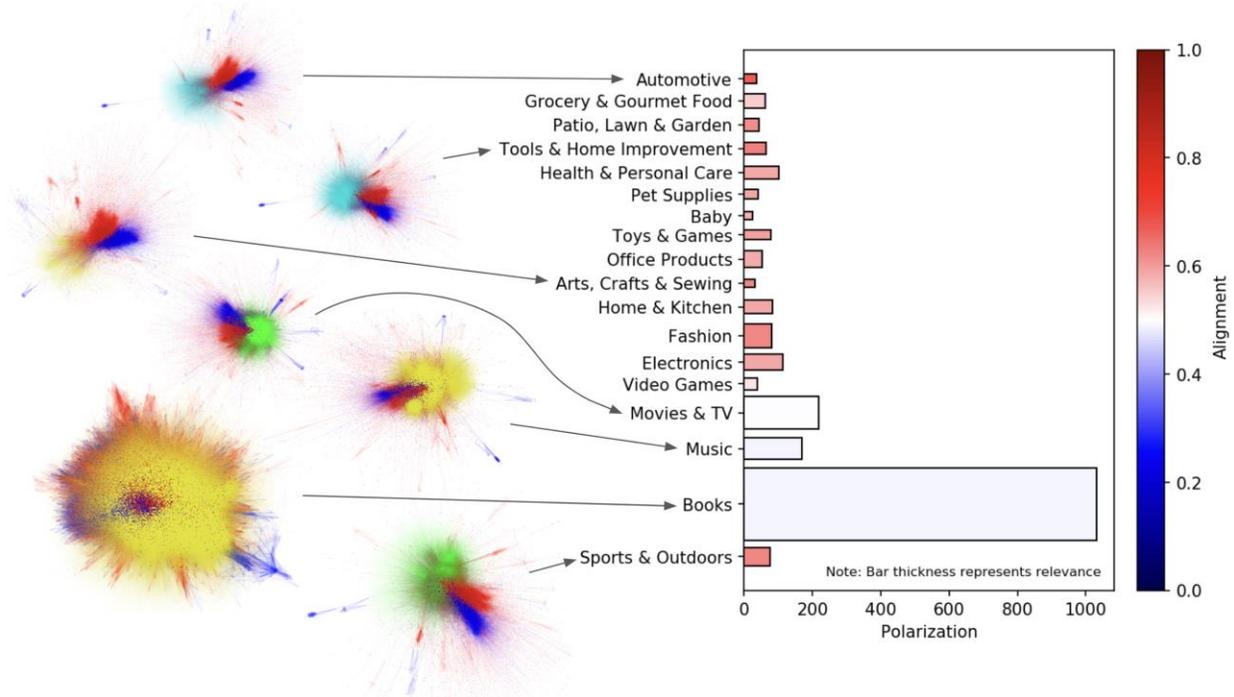

Note: Spring force-directed projections of select categories are shown on the left-hand side of the figure to qualitatively show the extent of political relevance, alignment, and polarization.

Of note here is the finding that most political polarization in markets is consolidated among culture and entertainment categories, which are also the most politically relevant categories. On the other hand, market segments that are most strongly aligned with particular ideologies are smaller and tend to be less relevant and polarized. In other words, though these categories have fewer edges to political products, the edges they do have to such products go predominantly to only one political ideology. This finding indicates that one may be able to better classify the political ideologies of a market segment through their consumption of products from the less relevant but high information (highly aligned) categories such as these.

More importantly, however, these results show how lifestyle politics are pervasive in markets. All categories had measurable political alignments. For example, music, movies, and TV may not be explicitly liberal in nature, nor are automotive products clearly conservative; however, the fact that the products in these categories are often co-purchased and co-reviewed with (or co- reviewed with products that tend to be co-purchased with – a 3$^{rd}$ order neighboring relation) explicitly political products provides sufficient information to classify them as being politically aligned and to detect the presence of lifestyle politics.

### 3.3 Moral Sentiment Classification

If lifestyle politics are pervasive in markets, what factors best explain their presence? Though the individual- and platform-level network dynamics examined so far may explain these results well, the lifestyle politics we observe among product categories that are not explicitly related to political products may also be explained by those products' inherent moral attributes, which tend to appeal to partisans differently. To test this possibility, tweet data from the morality dataset were randomly split into training (16,950) and evaluation sets (4,282) and used to create a language classification model that could classify the presence of multiple moral and non-moral labels in text. Before the multi-label classification model was trained, a pretrained RoBERTa language transformer model[11] (Liu et al. 2019) was fine-tuned for three epochs on the training set of tweets to better learn unique syntactical, grammatical, and other lexical features in the moral dataset's text that differ from text on which the model was originally trained.

Natural language processing models based on BERT (Devlin et al. 2019) often perform extremely well on classification tasks. This is because in contrast to previous language models that have objectives like learning to predict words that come before, after, and around a target word more globally – including word2vec (Mikolov et al. 2013) and GloVe (Pennington, Socher, and Manning 2014) – and in contrast to those that learn these dynamics in addition to how they vary across linguistic contexts – such as ELMo (Peters et al. 2018) and ULMFiT (Howard and Ruder 2018) – BERT learns to represent language by predicting self-masked sections of text (i.e., censoring words in texts and learning to predict them given the surrounding context) and by predicting whether sentences follow others (to infer context). The RoBERTa model chosen for this paper was selected because it extends BERT's unsupervised and deeply bidirectional approach by using ten times more training data, training on long sequences of text, using dynamic masking instead of static masking, and modifying the next sentence prediction objective to improve model performance.

As in the RGCN training pipeline, to enhance the RoBERTa model's training, automatic Bayesian hyperparameter was used to discover the best fitting model for multi-label classification of moral values in tweets. For this task, the Bayesian optimizer could choose to use the fine-tuned model or an untuned model in addition to modifying hyperparameters for learning rate, weight decay, Adam epsilon, and gradient norming/clipping value. The best-fitting model from this process was a fine-tuned model with default values for hyperparameters[12]. At testing, this model had a label ranking average precision (LRAP) score of 88.47%. This score represents the proportion of highly-ranked labels that are true labels, and it ranges from 0 – 100%. The binary cross entropy loss was 0.2775.

---

[11] This model is available from Hugging Face (https://huggingface.co/distilroberta-base) and was constructed, fine-tuned, and trained using the `simpletransformers` library in Python (https://simpletransformers.ai/docs/classification-models/#multilabelclassificationmodel).

[12] Learning rate = 4E-5, weight decay = 1E-6, Adam epsilon = 1E-8, maximum gradient normalization = 1.0.

The trained RoBERTa model was applied to the first sampling wave of Amazon reviews to predict the probability of each moral value (or non-morality) in each review's text. Several hundred predictions of moral values in reviews were randomly selected and manually evaluated to validate that the model's predictions made sense and had face validity. **Table 2** includes the number of reviews that had any probability of having a given label in addition to probability mass of each label across all reviews. Most review texts were primarily non-moral; however, a large number of reviews did express moral sentiment. Authority was most prevalent, which is not surprising given that most reviews are intended to state an opinion about a product's utility and recommend others to buy or avoid the product. It is also not surprising that the second most prevalent moral sentiment is harm, since negativity bias drives people to review more often if they do not like a product than if they like a product. Other moral sentiments are relatively similarly distributed across reviews.

Table 2: Presence and Probability Mass of Moral Sentiments across Review Texts

| Moral Sentiment | Presence (Probability Mass) |
|---:|:---:|
| *Care* | 76,183.03 (6.37%) |
| *Harm* | 123,330.05 (10.32%) |
| *Fairness* | 59,490.64 (4.96%) |
| *Cheating* | 84,065.39 (7.03%) |
| *Loyalty* | 97,226.66 (8.13%) |
| *Betrayal* | 85,122.68 (7.12%) |
| *Authority* | 178,451.31 (14.93%) |
| *Subversion* | 104,777.05 (8.77%) |
| *Purity* | 74,073.55 (6.20%) |
| *Degradation* | 77,224.51 (6.46%) |
| *Non-moral* | 1,154,487.88 (96.59%) |

Note: Presence is the number of reviews that had any probability of having a given label. The total number of reviews classified equaled 1,195,268.

## 3.4 Predicting Lifestyle Politics

To measure the degree of association between lifestyle politics, individual- and platform-network effects, and moral attributes, one must first measure the magnitude of lifestyle politics entwined with products. After merging the 20-core author-product heterogeneous network (3,647,388 nodes and 25,432,541 edges) with review data from first sample wave (2,319,244 rows of data, including reviewers' review data and product-level metadata) including labels for moral sentiments, the data remaining included 1,481,887 rows of product and author information from both datasets. Political relevance and alignment measures were extracted from reviewers by modifying Shi et al.'s (2017) equations (**Equation Set 1**) to focus on reviewers' edges to political and conservative products, respectively[13]. **Appendix A** lists and describes variables extracted from these steps in more detail.

The alignment of products' lifestyle politics is estimated using the RGCN from **Section 3.1** trained to classify products' political alignment, as this model implicitly learned to predict lifestyle politics. While iteratively training and hand-evaluating the RGCN, explicitly political products became less prevalent over time while implicitly political products (e.g., documentaries on climate change and gun rights t-shirts) became more prevalent, indicating that the model came to learn lifestyle politics through these not explicitly political products' direct and indirect relations with explicitly political products. This phenomenon is an example of machine learning bias, a well-documented outcome of training machine learning models on data that include dynamics that are associated with social and demographic factors that are not specified and well-controlled or when training data are not representative of all social groups and then do not generalize appropriately across them (Mehrabi et al. 2019). Examples of bias from previous research include racial bias in health care (Obermeyer et al. 2019), racial and geographic bias in policing and policy making (Courtland 2018), and gender bias in learning word representations – such as stereotypes like "man is to computer programmer as woman is to homemaker" (Bolukbasi et al. 2016; Gonen and Goldberg 2019).

---
[13] In the author context, political relevance measures the proportion of reviews authors make of political products, and political alignment measures the proportion of political product reviews that are of conservative products.

**Table 3: Stereotypical Political Relationships among Products Reflecting Lifestyle Politics**

| *"Liberal" Lifestyle Products* | *"Conservative" Lifestyle Products* |
|---|---|
| • Stainless steel juicer | • Grilling/BBQ equipment |
| • Organic cotton eco tote | • Leather pistol holster |
| • Purple cardigan | • Carhartt Jacket |
| • iPod speakers | • Bushnell hunting field tripod |
| • Bruce Springsteen | • Walker Texas Ranger |
| • Ecological aesthetics | • Colonial Williamsburg gardening |
| • Swarovski dog collar | • Rottweiler wire basket muzzle |
| • "Inspiration for Life" | • "Southern Literature" |
| • "Nuclear Disaster" | • "Fight Against Radical Islam" |
| • "Understanding Abortion" | • "Pro-Life Reflections" |

The RGCN learned to predict stereotypical lifestyle politics alignments among products that reflect themes such as those shown in **Table 3**. Liberals, for example, were predicted to be organic, eco-friendly consumers who make their own juice, oppose nuclear power/weapons, and support pro-choice rights. By contrast, the RGCN predicted that conservatives are blue collar consumers who like to hunt, grill, and be informed about Southern lifestyles, radical Islam, and pro-life apologetics. These trends are similar to those found in earlier studies of lifestyle politics (DellaPosta et al. 2015; Y. Shi et al. 2017) that found empirical support for liberals' affinities for lattes and conservatives' support for bird hunting.

Given these results, a lifestyle politics alignment variable was created for every product that was not hand-labeled or labeled using the RGCN by taking the logged odds of the probability that a product is conservative compared to liberal and clipping the values between [0.0001, 0.9999]. This logistic transform linearizes the probabilistic sigmoid function that represents products' alignment prediction from the RGCN[14]. These values were min-max scaled from approximately -10 to 10 to 0 to 1, with 0 indicating liberal lifestyle politics and 1 indicating conservative lifestyle politics, to improve their interpretation as going from completely aligned with liberal lifestyles to completely aligned with conservative lifestyles. Evaluating the distribution of this variable over each of the product categories revealed that lifestyle politics have the strongest entwinement with cultural and entertainment products[15]. After dropping products in categories less strongly affected by lifestyle politics, 1,006,172 rows of data remained. These data were consolidated further by removing data from reviewers who had not posted at least five reviews, leaving 149,813 rows of data[16]. Many of these covariates were not normally distributed, so they were Yeo-Johnson power transformed and standard-normalized, after which Shapiro-Wilks tests confirmed the covariates' distributions were substantially closer to a normal distribution[17].

---

[14] If lifestyle politics alignment is not made linear with the logistic transform, the resulting mixed effects model is unstable and has residuals that are not normally distributed around 0.

[15] Attempting to model lifestyle politics with data from other categories would not allow the model to converge.

To determine which covariates best predict products' lifestyle alignment, a beta generalized linear mixed effects regression model was constructed using R's `GLMMadaptive` library. This model is most appropriate for the given data as the dependent variable is not continuous but ranges from 0 to 1 and since data are complexly structured with repeated measures existing for reviewers through their multiple reviews of different products. The main covariates of interest for this model include product-level political alignment and relevance, reviewer-level political alignment and relevance, and review-level moral sentiments. Several statistical interactions were also included to test for moderation effects between these predictors (e.g., relevance and alignment). Other covariates are modeled for statistical controls (e.g., reviews' average helpfulness score, reviews' overall rating score for a product, and products' category). Random effects are included for reviewers[18].

**Figure 4** presents the coefficients of the best-fitting model for predicting the alignment of lifestyle politics among products[19] (over 25 configurations of linear and mixed effects models). Excluding category coefficients, which are binary indicators, one can approximately interpret coefficients as the amount of change in lifestyle politics alignment that is associated with a one standard deviation change in the given covariate[20]. These results reveal that substantial category-level differences in lifestyle politics exist, which aligns with the results and insights drawn from **Figure 3**. Books and Movies & TV have significant liberal alignment with lifestyle politics, whereas videogames have significant conservative liberal alignment. Interestingly, while the Sports & Outdoors category has substantial conservative political alignment, in the mixed effects model it is significantly aligned with liberal lifestyles. This inconsistency could be because less active reviewers were not included in the lifestyle politics model, nor were short reviews, and liberal reviewers tended to write more and longer reviews.

Product-level political alignment has the greatest association with product-level lifestyle politics. An approximate interpretation of this coefficient is that products that have a political alignment one standard deviation above the mean (i.e., a standard deviation conservative lean) have a 75.67% probability of having conservatively aligned lifestyle politics (i.e., leaning conservatively 0.2567 past the 0.5 neutral point), on average[21]. This relation is compounded through a positive interaction with product-level political relevance, indicating that products that are both politically relevant and politically aligned are especially more likely to have strong lifestyle political alignment. For example, products that are co-purchased or co-reviewed often with explicitly political products, where those products tend to be highly liberal in political alignment, also have stronger liberal alignments in lifestyle politics on average. Author-level political alignment and political relevance also have statistically significant associations with lifestyle politics. The coefficient for author-level political alignment (0.20) is almost the same as the product-level political relevance's main effect (0.21); however, overall, the coefficients for the product-level covariates are substantially larger than their author-level analogs, giving evidence to the argument that product- and platform-network effects have a stronger association with lifestyle politics than author-level network effects.

---

[16] If low activity reviewers are not removed, the mixed effects model failed to converge due to segmentation errors from attempting to properly generate random effects for reviewers who had one or very few reviews.

[17] Operations were performed with the `stats.yeojohnson` and `stats.shapiro` functions of Python's `scipy` package (Virtanen et al. 2020). If covariates are not standardized and transformed, the model will not converge.

[18] The `GLMMadaptive` library only allows for one grouping random effect, which is why "main" categories are included as indicator variables and not modeled as random effects.

[19] AIC/BIC tests determined this model was the best-fitting among 25 configurations of linear and mixed effects models that predicted lifestyle politics alignment among products. Modeling lifestyle politics on its original logit scale of -10 to 10 with a linear mixed effects model converged; however, residuals were not normally distributed. The coefficients were in the same direction and of the same magnitude as those in **Figure 3**, though.

[20] This is only approximately correct, as covariates were power transformed before standard-normalizing.

**Figure 4: Beta Linear Mixed Effects Regression Model Results Predicting Lifestyle Politics**

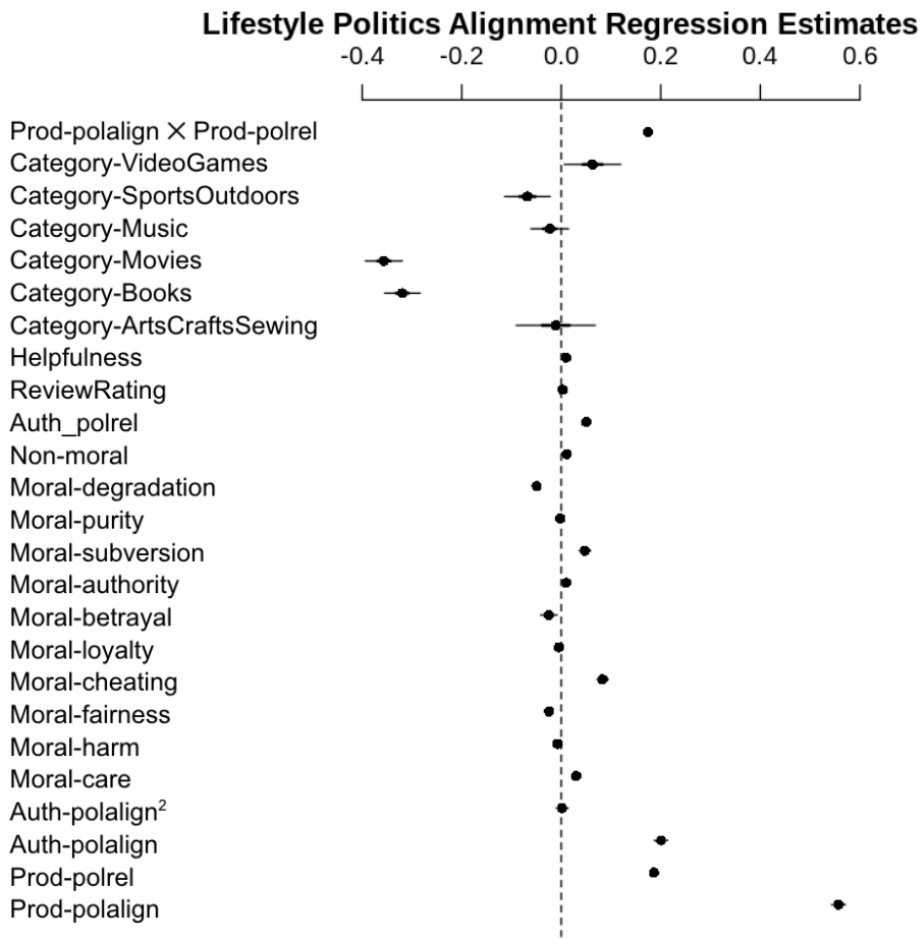

Note: Lifestyle politics ranges from 0 to 1, where 0 is liberal alignment and 1 is conservative.

Among the moral sentiments, most values have small but statistically significant associations with product-level lifestyle politics. Lifestyle politics' positive correlations with subversion, authority, and cheating and its negative correlations with harm and fairness align well with the associations predicted by moral foundations theory, which argues that these moral values often appeal more to conservatives than liberals (Graham et al. 2009). Lifestyle politics' positive correlation with care and its negative correlations with degradation and betrayal, on the other hand, are reverse of what is expected by moral foundations theory, which often finds that liberals give more weight to issues of care/harm than conservatives and that conservatives usually emphasize degradation and betrayal more than liberals. That being the case, however, Koleva and colleagues (2012) show that in many cases of "cultural war attitudes," the alignments between moral sentiments and partisan topics are often mixed from liberals' and conservatives' general support of the core moral foundations. The results in **Figure 4**, therefore, provide additional support for their argument. In the context of this paper, though, these results demonstrate that products' moral attributes are associated with lifestyle politics; however, the magnitude of these relationships is substantially less than that of product- or author-level political characteristics.

---

[21] This probability is derived by first converting the min-max scaled lifestyle politics coefficient to its original logit form: (0.5572 * (10 – -10)) + -10 = 1.144. Then, convert this logit to a probability: e(1.144)/(1 + e(1.144)) = 0.7567.

To determine how much variance in lifestyle politics alignment is explained by these covariates, a linear mixed effects regression model[22] was created that was identical to the beta mixed effects model presented in **Figure 4** in all ways except that the linear version could properly model random effects for reviewers and "main" categories as well as crossed random effects for reviewers nested within categories (through multiple reviews of products in the same category). The marginal $R^2$ of the model, which represents how much of the model variance is explained by the fixed effects only, equaled 0.3938. The adjusted $R^2$ of the model, which represents how much of the model variance is explained by the complete model (including fixed, random, and residual effects), equaled 0.6866. In other words, the product, reviewer, morality, and category level covariates presented in **Figure 4** explain almost 70% of the variance in products' lifestyle politics.

## 4  Conclusion

This study examined whether market networks are politically polarized, if these political dynamics crossover to non-political market segments to generate lifestyle politics, and if lifestyle politics are better explained by moral values (e.g., inherent qualities of products that appeal to partisans' social and cognitive differences) versus individual-level (e.g., selection effects from authors' affinities for politically relevant and aligned products) or product-level network dynamics (e.g., products often co-purchased or co-reviewed with political products may spawn filter effects as platforms' recommendation engines learn to suggest the political products to other potential consumers of the non-political products). Bayesian estimates of political relevance, alignment, and polarization find cultural segments are over four times more polarized than the second most polarized segment. The extent of political polarization in other segments is relatively less; however, even small categories like automotive parts have notable political alignment indicative of lifestyle politics. These results indicate that lifestyle politics spread deep and wide across markets – extending prior work on the partisan divide in scientific markets (F. Shi et al. 2017).

Secondly, this study demonstrated that semi-supervised graph neural networks implicitly learn to classify lifestyle politics similarly to how word embedding models learn gender biases (Bolukbasi et al. 2016). With these results, beta mixed effects regression models find product-level political alignment and relevance are more strongly associated with lifestyle politics than their author-level equivalents (2.75 – 3.7 ✕ difference, respectively). Moral sentiments, by contrast, are significantly associated with lifestyle politics in many ways that are predicted by moral foundations theory, but in other ways the observed relationships go against the theory's predictions. Overall, though, these covariates' effects were small compared to the product- and author-level covariates. Finally, even with these covariates, market segments' differences in lifestyle politics persisted, and the complete model failed to explain about 30% of the variance in lifestyle politics. Therefore, other variables are still missing which could help us better understand the growth and spread of lifestyle politics.

---

[22] R's `lmer` function from the `lme4` package (https://cran.r-project.org/web/packages/lme4/index.html) was used for this task because the `GLMMAdaptive` package could not produce the necessary information for these calculations.

Though it was beyond the scope of this study, our knowledge of lifestyle politics may be expanded by analyzing how lifestyle politics and the social dynamics associated with it evolve over time and across products and the individuals who review them. For example, Baldassarri and Park (2020) use longitudinal data to show how while partisans' support for different moral and political issues is divided, the two groups' opinions have actually been converging over time, with conservatives adopting more secular values more quickly than liberals over time (even though they are still less secular than liberals at the present time). It could also be the case that in market networks political polarization and lifestyle politics are evolving at different rates across partisan groups over time. Also, the fact that product-level political alignment has a stronger association with lifestyle politics than author-level political alignment could reflect a similar process to what Macy and colleagues' (2019) found in their work on opinion cascades, where the initial alignment of topics is most salient for signaling and attracting like-minded supporters (and attracting opposition from detractors). By analyzing the temporal dynamics of lifestyle politics, future research could try to better decompose how product- and author-level political affinities accumulate and shape one another over time.

Finally, while this study used graph and language data together to discover latent variables within the markets' products and consumers, the use of these two data types could be integrated further by better modeling the fact that polarization is more likely when liberals review liberal products favorably and review conservative products negatively (analogously for conservatives). This treats language data as an edge feature that contributes affinity/aversion information to the relationships between products and reviewers, with the objective of better measuring polarization as walled-off clustering with the presence of negative ties between groups and positive ties within groups (e.g., Leifeld 2017). Similarly, in their work on partisan polarization in discourse on shootings, Demszky et al. (2019) find partisans are polarized in how they discuss law & policy, identity & ideology, solidarity, remembrance and other topics over time.

## Acknowledgements

This work was supported in part by NSF grant SES-1756822. The views and conclusions presented in this study are those of the author and should not be interpreted as necessarily representing the official policies or endorsements – either expressed or implied – of NSF, the U.S. Government, Cornell University, or Graphika Inc. The author wishes to thank Megan Zhou for assistance with sampling Amazon network data and evaluating sample results. The author also wishes to express gratitude to Cornell University's Social Dynamics Laboratory and Cornell University's class on "Culture Wars and Polarization" for comments and feedback on this paper. Much Gratitude is also extended to Stephen Parry of the Cornell Statistical Consulting Unit for consulting on various parts of the mixed effects modeling for this study.

*Proceedings of NAACL-HLT 2019*.

Graham, Jesse, Jonathan Haidt, Sena Koleva, Matt Motyl, Ravi Iyer, Sean P. Wojcik, and Peter H. Ditto. 2013. *Moral Foundations Theory: The Pragmatic Validity of Moral Pluralism*. Vol. 47. 1st ed. Copyright © 2013, Elsevier Inc. All rights reserved.

Graham, Jesse, Jonathan Haidt, and Brian A. Nosek. 2009. "Liberals and Conservatives Rely on Different Sets of Moral Foundations." *Journal of Personality and Social Psychology* 96(5):1029–46.

He, Ruining and Julian Mcauley. 2016. "Ups and Downs: Modeling the Visual Evolution of Fashion Trends with One-Class Collaborative Filtering." *WWW* 1–11.

Hoover, Joe, Gwenyth Portillo-Wightman, Leigh Yeh, Shreya Havaldar, Aida Mostafazadeh Davani, Ying Lin, Brendan Kennedy, Mohammad Atari, Zahra Kamel, Madelyn Mendlen, Gabriela Moreno, Christina Park, Tingyee E. Chang, Jenna Chin, Christian Leong, Jun Yen Leung, Arineh Mirinjian, and Morteza Dehghani. 2020. "Moral Foundations Twitter Corpus: A Collection of 35k Tweets Annotated for Moral Sentiment." *Social Psychological and Personality Science* 11(8):1057–71.

Howard, Jeremy and Sebastian Ruder. 2018. "Universal Language Model Fine-Tuning for Text Classification." Pp. 328–39 in *ACL 2018 - 56th Annual Meeting of the Association for Computational Linguistics, Proceedings of the Conference (Long Papers)*. Vol. 1.

Iyengar, Shanto, Yphtach Lelkes, Matthew Levendusky, Neil Malhotra, and Sean J. Westwood. 2019. "The Origins and Consequences of Affective Polarization in the United States." *Annual Review of Political Science* 22:129–46.

Iyengar, Shanto, Gaurav Sood, and Yphtach Lelkes. 2012. "Affect, Not Ideology: A Social Identity Perspective on Polarization." *Public Opinion Quarterly* 76(3):405–31.

Kapner, Suzanne and Dante Chinni. 2019. "Are Your Jeans Red or Blue? Shopping America's Partisan Divide." *The Wall Street Journal*.

Kipf, Thomas N. and Max Welling. 2017. "Semi-Supervised Classification with Graph Convolutional Networks." *ICLR* 1–14.

Koleva, Spassena P., Jesse Graham, Ravi Iyer, Peter H. Ditto, and Jonathan Haidt. 2012. "Tracing the Threads: How Five Moral Concerns (Especially Purity) Help Explain Culture
19

Zheng, Da, Minjie Wang, Quan Gan, Zheng Zhang, and Geroge Karypis. 2020. "Scalable Graph Neural Networks with Deep Graph Library." Pp. 3521–22 in *Proceedings of the ACM SIGKDD International Conference on Knowledge Discovery and Data Mining*. New York, NY, USA: Association for Computing Machinery.

# APPENDIX

## A  Information Extracted from Political, Market, and Morality Samples

| *Sample* | *Variable (type)* | *Description* |
|---|---|---|
| *Political Products* | Name (node) | Product's name |
| *Political Products* | Conservative (node) | Is a product conservative? |
| *Political Products* | Liberal (node) | Is a product liberal? |
| *Market Network* | Author (node) | Is a node an author? |
| *Market Network* | Product (node) | Is a node a product? |
| *Market Network* | Brand (node) | What brand is the product? |
| *Market Network* | Categories (node) | Product's category membership |
| *Market Network* | Conservative (node) | Is a product conservative? |
| *Market Network* | Liberal (node) | Is a product liberal? |
| *Market Network* | Id (node) | ASIN for products, ReviewerId for reviewers |
| *Market Network* | Name (node) | Product's name, reviewer's username |
| *Market Network* | RatingAv (node) | Average review rating for product |
| *Market Network* | Rank (node) | Product ranking across categories |
| *Market Network* | Sample (node) | In what wave was the product/author sampled |
| *Market Network* | Rating (edge) | Reviewer's rating of product |
| *Market Network* | Helpfulness (edge) | How helpful was reviewer's review |
| *Market Network* | Author-Product (edge) | Edge going from author to product |
| *Market Network* | Product-Product (edge) | Edge going from product to product |
| *Market Network* | Review time (edge) | Unix time of review creation |
| *Morality* | Care (edge) | Care moral foundation (opposite of harm) |
| *Morality* | Harm (edge) | Harm moral foundation (opposite of care) |
| *Morality* | Fairness (edge) | Fairness moral foundation (opposite of cheating) |
| *Morality* | Cheating (edge) | Cheating moral foundation (opposite of fairness) |
| *Morality* | Loyalty (edge) | Loyalty moral foundation (opposite of betrayal) |
| *Morality* | Betrayal (edge) | Betrayal moral foundation (opposite of loyalty) |
| *Morality* | Authority (edge) | Authority moral foundation (opposite of subversion) |
| *Morality* | Subversion (edge) | Subversion moral foundation (opposite of authority) |
| *Morality* | Purity (edge) | Purity moral foundation (opposite of degradation) |
| *Morality* | Degradation (edge) | Degradation moral foundation (opposite of purity) |
| *Morality* | Non-moral (edge) | Review text does not present with morally-laden language |

# B  Degree Distribution by Edge Type

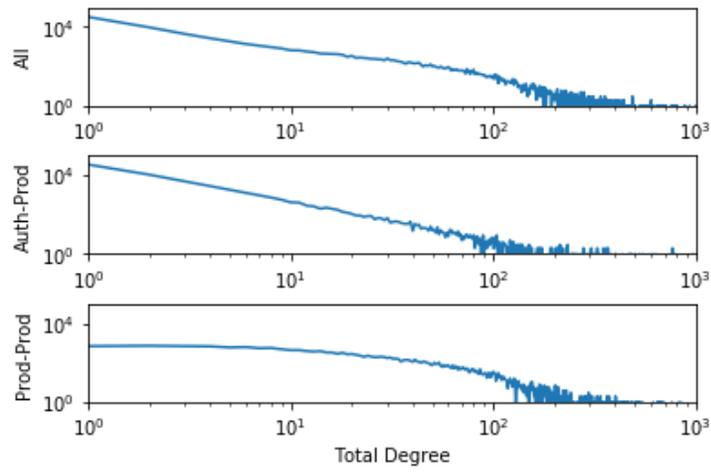

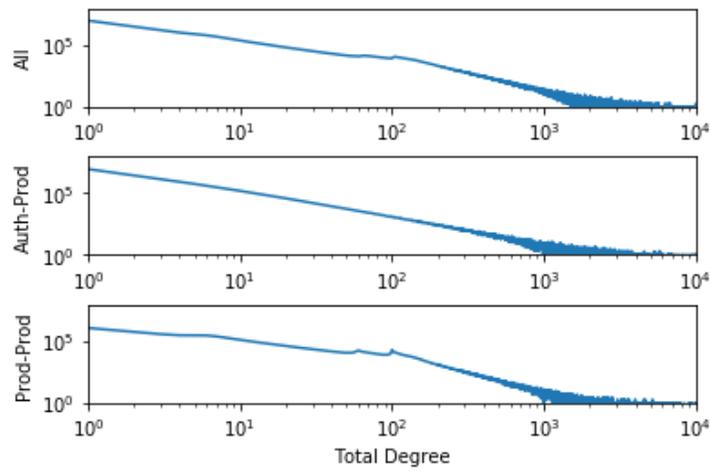

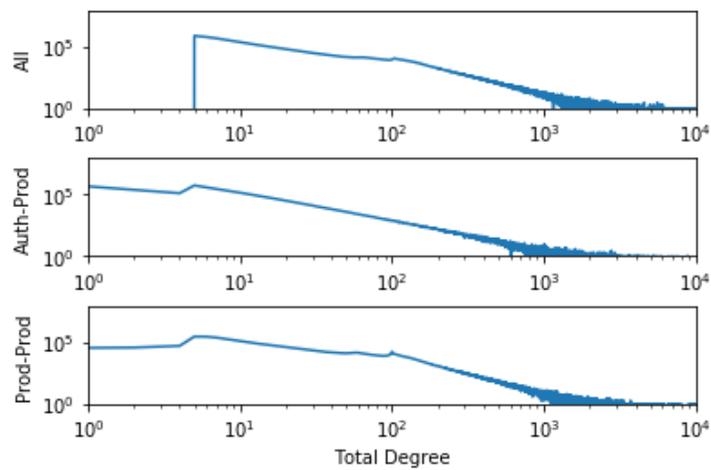

## C Decision Thresholds versus Accepted Label Proportions for RGCNs

### First RGCN Application (Classifications: liberal, conservative)

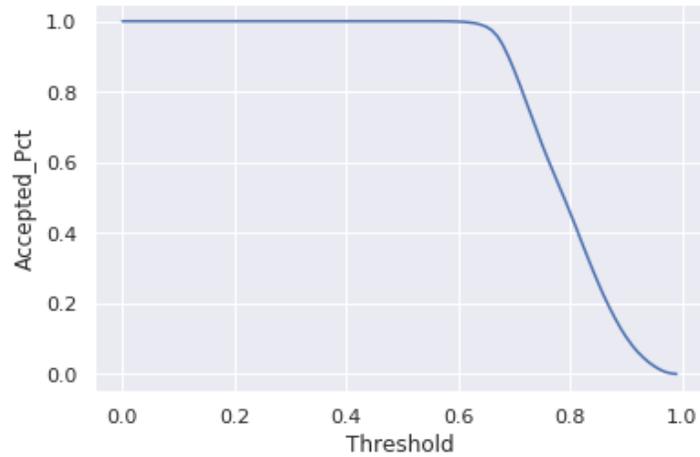

### Second RGCN Application (Classifications: liberal, conservative, non-political)

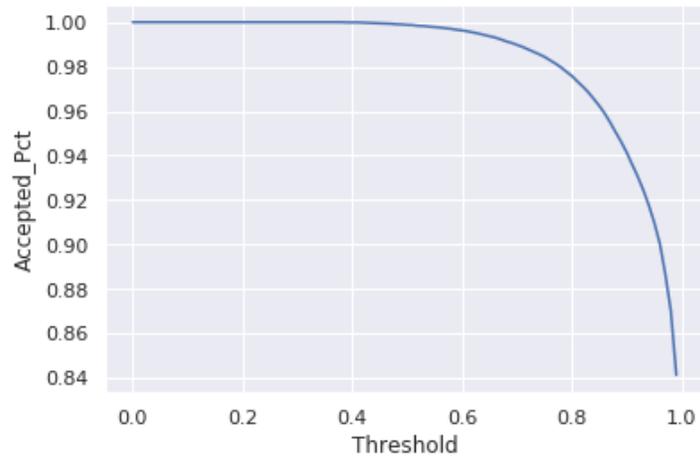

### Fifth/Final RGCN Application (Classifications: liberal, conservative, non-political)

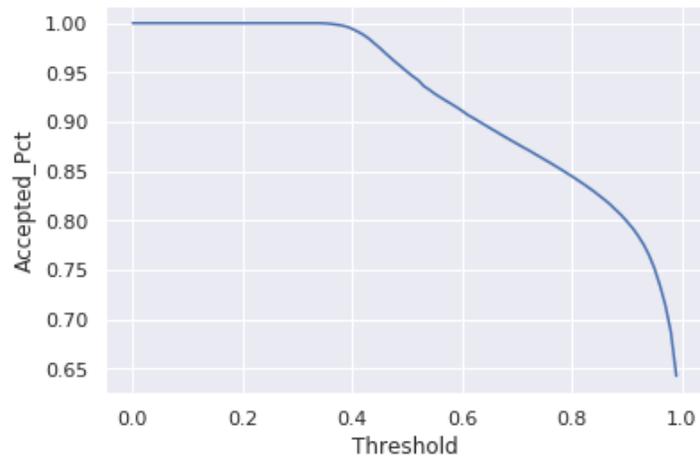

# D  Products in Categories and Brands

The original market dataset included dozens of nested categories that were collapsed into smaller "main" and "big" categories. For example, in the "main" category, "CDs & Vinyl" was merged into the "Music" category. Then, in the "big" category, the "Music" was merged into "Culture."

| **5-core Heterogeneous Graph** | **20-core Heterogeneous Graph** |
|---|---|
| *"Main" categories: n = 18* | *"Main" categories: n = 18* |
| o   Books: 927,006 | o   Sports & Outdoors: 59,109 |
| o   Sports & Outdoors: 133,156 | o   Books: 513,419 |
| o   Music: 287,844 | o   Music: 149,700 |
| o   Movies & TV: 96,134 | o   Movies & TV: 61,665 |
| o   Video Games: 27,448 | o   Video Games: 18,287 |
| o   Electronics: 274,776 | o   Electronics: 124,167 |
| o   Fashion: 115,537 | o   Fashion: 50,464 |
| o   Tools & Home Improvement: 137,206 | o   Home & Kitchen: 78,567 |
| o   Home & Kitchen: 177,057 | o   Arts, Crafts & Sewing: 23,612 |
| o   Arts, Crafts & Sewing: 49,066 | o   Office Products: 28,718 |
| o   Office Products: 63,003 | o   Toys & Games: 90,630 |
| o   Toys & Games: 167,205 | o   Baby: 14,198 |
| o   Baby: 24,375 | o   Pet Supplies: 24,392 |
| o   Pet Supplies: 50,480 | o   Health & Personal Care: 115,447 |
| o   Health & Personal Care: 241,510, | o   Tools & Home Improvement: 55,097 |
| o   Patio, Lawn & Garden: 59,488, | o   Patio, Lawn & Garden: 28,739 |
| o   Grocery & Gourmet Food: 65,826, | o   Grocery & Gourmet Food: 35,806 |
| o   Automotive: 129,902 | o   Automotive: 38,204 |
| *"Big" categories: n = 5* | *"Big" categories: n = 5* |
| o   Culture: 1263,916, | o   Entertainment: 229,691 |
| o   Entertainment: 423,943, | o   Culture: 686,731 |
| o   Products: 453,316, | o   Products: 203,349 |
| o   Home: 569,479, | o   Home: 236,413 |
| o   Personal & Family: 316,365 | o   Personal & Family: 154,037 |

```
main_cats_regroup_big = {
  "Books":"Culture",
  "Music":"Culture",
  "Arts, Crafts & Sewing":"Culture",
  "Movies & TV":"Entertainment",
  "Video Games":"Entertainment",
  "Toys & Games":"Entertainment",
  "Sports & Outdoors":"Entertainment",
  "Health & Personal Care":"Personal & Family",
  "Baby":"Personal & Family",
  "Pet Supplies":"Personal & Family",
  "Electronics":"Products",
  "Fashion":"Products",
  "Office Products":"Products",
  "Patio, Lawn & Garden":"Home",
  "Home & Kitchen":"Home",
  "Tools & Home Improvement":"Home",
  "Grocery & Gourmet Food":"Home",
  "Automotive":"Home"
}
```